\documentclass[a4paper,11pt]{article}
\pdfoutput=1 

\usepackage{jheppub} 

\usepackage[T1]{fontenc} 

\usepackage{amsthm}

\usepackage{slashed} 

\title{\boldmath Curvature constraints in heterotic Landau-Ginzburg models}

\author{Richard S. Garavuso}

\affiliation{Physical Sciences, Kingsborough Community College, The City University of New York,\\ 2001 Oriental Boulevard, Brooklyn, NY 11235-2398, USA}

\emailAdd{richard.garavuso@kbcc.cuny.edu}

\abstract{
In this paper, we study a class of heterotic Landau-Ginzburg models.
We show that the action can be written as a sum of BRST-exact and non-exact terms.
The non-exact terms involve the pullback of the complexified K{\"a}hler form to the worldsheet and terms arising from the superpotential, which is a Grassmann-odd holomorphic function of the superfields.
We then demonstrate that the action is invariant on-shell under supersymmetry transformations up to a total derivative.
Finally, we extend the analysis to the case in which the superpotential is not holomorphic.
In this case, we find that supersymmetry imposes a constraint which relates the nonholomorphic parameters of the superpotential to the Hermitian curvature.
Various special cases of this constraint have previously been used to establish
properties of Mathai-Quillen form analogues which arise in the corresponding heterotic Landau-Ginzburg models.
There, it was claimed that supersymmetry imposes those constraints.
Our goal in this paper is to support that claim.
The analysis for the nonholomorphic case also reveals a constraint imposed by supersymmetry that we did not anticipate from studies of Mathai-Quillen form analogues.
        }

\begin{document} 
\maketitle
\flushbottom
\numberwithin{equation}{section}
\newtheorem{theorem}{Theorem}[section]
\newtheorem{lemma}[theorem]{Lemma}
\newtheorem{proposition}[theorem]{Proposition}
\newtheorem{corollary}[theorem]{Corollary}
\newtheorem{definition}[theorem]{Definition}
\newtheorem{remark}[theorem]{Remark}
\section{\label{Intro}Introduction}

A Landau-Ginzburg model is a nonlinear sigma model with a superpotential.
For a \emph{heterotic} Landau-Ginzburg model
\cite{Witten:Phases, DistlerKachru:0-2-Landau-Ginzburg, AdamsBasuSethi:0-2-Duality, MelnikovSethi:Half-twisted, GuffinSharpe:A-twistedheterotic, MelnikovSethiSharpe:Recent-Developments, GaravusoSharpe:Analogues}, 
the nonlinear sigma model possesses only
$ (0, 2) $
supersymmetry and the superpotential is a Grassmann-odd function of the superfields which may or may not be holomorphic.
These heterotic models have field content consisting of
$ (0, 2) $ 
bosonic chiral superfields
$ \Phi^i = (\phi^i, \psi^i_+ )$
and
$ (0, 2) $ 
fermionic chiral superfields 
$ 
  \Lambda^a 
    = \left( \lambda^a_-, 
             H^a, 
             E^a
      \right) 
$,
along with their conjugate antichiral superfields
$ 
  \Phi^{\overline{\imath}} 
    = \left(
             \phi^{\overline{\imath}}, 
             \psi^{\overline{\imath}}_+ 
      \right)
$
and
$ 
  \Lambda^{\overline{a}} 
    = \left( 
             \lambda^{\overline{a}}_-, 
             \overline{H}^{\overline{a}}, 
             \overline{E}^{\overline{a}}
      \right) 
$.
The 
$ \phi^i $ 
are local complex coordinates on a K{\"a}hler manifold 
$ X $. 
The 
$ E^a $
are local smooth sections of a Hermitian vector bundle 
$ \mathcal{E} $
over 
$ X $,
i.e. 
$ E^a \in \Gamma(X, \mathcal{E}) $. 
The
$ H^a $
are nonpropagating auxiliary fields.
The fermions couple to bundles as follows:
\begin{align*}
\psi^i_+ 
  \in \Gamma
      \left(
             K^{1/2}_{\Sigma}
             \otimes
             \Phi^* \!
             \left(
                    T^{1,0} 
                    X
             \right)      
      \right),
\qquad
&
\lambda^a_- 
  \in \Gamma
      \left(
             \overline{K}^{1/2}_{\Sigma}
             \otimes
             \left(
                    \Phi^* 
                    \overline{\mathcal{E}}
             \right)^{\vee}
      \right),
\\
\psi^{\overline{\imath}}_+
  \in \Gamma
      \left(
             K^{1/2}_{\Sigma}
             \otimes
             \left(
                    \Phi^* \!
                    \left(
                           T^{1,0}
                           X
                    \right)       
             \right)^{\vee}
      \right),
\qquad
&
\lambda^{\overline{a}}_-
  \in \Gamma
      \left(
             \overline{K}^{1/2}_{\Sigma}
             \otimes
             \Phi^* 
             \overline{\mathcal{E}}
      \right),                    
\end{align*}
where
$ \Phi: \Sigma \rightarrow X $
and
$ K_{\Sigma} $
is the canonical bundle on the worldsheet
$ \Sigma $.
In 
\cite{GuffinSharpe:A-twistedheterotic},
heterotic Landau-Ginzburg models with superpotential of the form
\begin{equation}
  W = \Lambda^a \,
      F_a \, ,
\end{equation}      
where 
$ 
  F_a 
  \in
  \Gamma
  \left(
         X, 
         \mathcal{E}^{\vee}  
  \right)
$,
were considered.
In this paper, we will study supersymmetry in these heterotic Landau-Ginzburg models with 
$ E^a = 0 $.
Such models yield the
$ (2,2) $ 
Landau-Ginzburg models of
\cite{GuffinSharpe:A-twisted}
when
$ \mathcal{E} = TX $ 
and
$ \Lambda^i \, F_i = \Lambda^i \, \partial_i W^{(2,2)} $,
where
$ W^{(2,2)} $
is the (2,2) superpotential.
It was claimed in 
\cite{GaravusoSharpe:Analogues}
that, when the superpotential is not holomorphic, supersymmetry imposes a constraint which relates the nonholomorphic parameters of the superpotential to the Hermitian curvature.
Our goal in this paper is to support that claim. 

This paper is organized as follows:
In section 
\ref{section:Action}, 
we will write down the action for the class of heterotic Landau-Ginzburg models that we are considering.
In section
\ref{section:Supersymmetry-holomorphic}, for the case of a holomorphic superpotential, we will show that the action can be written as a sum of
BRST-exact and non-exact terms.
We will then demonstrate that the action is invariant on-shell under supersymmetry transformations up to a total derivative.
Finally, in section
\ref{section:Supersymmetry-nonholomorphic},
we will extend the analysis to the case in which the superpotential is not holomorphic.
In this case, we will show that supersymmetry imposes a constraint which relates the nonholomorphic parameters of the superpotential to the Hermitian curvature.
The analysis for this case will also reveal another constraint imposed by supersymmetry.
\section{\label{section:Action}Action}

Let
$ X $ 
be a K{\"a}hler manifold with metric
$ g $, 
antisymmetric tensor
$ B $,
local real coordinates
$ \phi^{\mu} $, 
and local complex coordinates 
$ \phi^i $
with complex conjugates
$ \phi^{\overline{\imath}} $.
Furthermore, let 
$ \mathcal{E} $
be a vector bundle over
$ X $
with Hermitian fiber metric
$ h $.
We consider the action
\cite{GuffinSharpe:A-twistedheterotic}
of a heterotic Landau-Ginzburg model on
$ X $ 
with gauge bundle 
$ \mathcal{E} $:
\begin{align}
\label{action}
S &= 2t \int_{\Sigma} d^2 z 
     \left[ 
            \frac{1}{2} 
            \left( 
                   g_{\mu \nu} + i B_{\mu \nu}
            \right)
            \partial_z 
            \phi^{\mu} 
            \partial_{\overline{z}} 
            \phi^{\nu}
          + i 
            g_{\overline{\imath} i}
            \psi_+^{\overline{\imath}}
            \overline{D}_{\overline{z}} \psi_+^i
          + i 
            h_{a \overline{a}} 
            \lambda_-^a
            D_z \lambda_-^{\overline{a}}
     \right. 
\nonumber
\\
  &\phantom{= 2t \int_{\Sigma} d^2 z \left[ \right.}
   +
     \biggl. 
     F_{i \overline{\imath} a \overline{a}} \,
     \psi_+^i
     \psi_+^{\overline{\imath}}
     \lambda_-^a
     \lambda_-^{\overline{a}}
   + h^{a \overline{a}} 
     F_{a} 
     \overline{F}_{\overline{a}}
   + \psi_+^i 
     \lambda_-^{a} 
     D_i F_a 
   + \psi_+^{\overline{\imath}}
     \lambda_-^{\overline{a}} 
     \overline{D}_{\overline{\imath}} \overline{F}_{\overline{a}}
     \biggr].     
\end{align}
Here,
$ t $ 
is a coupling constant,
$ \Sigma $
is a Riemann surface, 
$ 
  F_a  
  \in 
  \Gamma
  \left(
         X, 
         \mathcal{E}^{\vee}
  \right) 
$,
and
\begin{align*}
\overline{D}_{\overline{z}} \,
\psi^i_+
   &= \overline{\partial}_{\overline{z}} \,
      \psi_+^i
    + \overline{\partial}_{\overline{z}} \,
      \phi^j \,
      \Gamma^i_{jk}
      \psi^k_+ \, ,
&D_z \lambda^{\overline{a}}_-
  &= \partial_z 
     \lambda_-^{\overline{a}}
   + \partial_z \phi^{\overline{\imath}}
     A^{\overline{a}}_{\overline{\imath} \overline{b}} \,
     \lambda^{\overline{b}}_- \, ,
\\
D_i F_a
  &= \partial_i F_a
   - A^{b}_{i a}
     F_b \, ,
& \overline{D}_{\overline{\imath}}
  \overline{F}_{\overline{a}}
  &= \overline{\partial}_{\overline{\imath}} \,
     \overline{F}_{\overline{a}}
   - A^{\overline{b}}_{\overline{\imath} \, \overline{a}} \,
     \overline{F}_{\overline{b}} \, ,
\\
A^b_{i a} 
  &= h^{b \overline{b}} \,
     h_{\overline{b} a, i} \, ,
& A^{\overline{b}}_{\overline{\imath} \, \overline{a}} 
  &= h^{\overline{b} b} \, 
     h_{b \overline{a}, \overline{\imath}} \, ,     
\\
\Gamma^i_{jk}
  &= g^{i \overline{\imath}} \,
     g_{\overline{\imath} k, j} \, ,         
& F_{i \overline{\imath} a \overline{a}}  
   &= h_{a \overline{b}} \,
      A^{\overline{b}}_{\overline{\imath} \, \overline{a}, i} \, .
\end{align*}
The fermions couple to bundles in the manner described in section
\ref{Intro}.

The action 
\eqref{action}
is invariant on-shell under the supersymmetry transformations
\begin{equation}
\begin{aligned}
\delta \phi^i
  &= i 
     \alpha_-
     \psi^i_+ \, ,
\\
\delta \phi^{\overline{\imath}}
  &= i 
     \tilde{\alpha}_-
     \psi^{\overline{\imath}}_+ \, ,          
\\
\delta \psi^i_+ 
  &= 
   - \tilde{\alpha}_-
     \overline{\partial}_{\overline{z}} \phi^i \, ,
\\
\delta \psi^{\overline{\imath}}_+
  &=
   - \alpha_-
     \partial_z \phi^{\overline{\imath}} \, ,
\\
\delta \lambda^{a}_-
  &= 
   - i
     \alpha_-
     \psi^j_+ \,
     A^a_{j b} \,
     \lambda^{b}_- 
   + i
     \alpha_-
     h^{a \overline{a}} \,
     \overline{F}_{\overline{a}} \, ,
\\
\delta \lambda^{\overline{a}}_-
  &= 
    - i
     \tilde{\alpha}_-
     \psi^{\overline{\jmath}}_+ \,
     A^{\overline{a}}_{\overline{\jmath} \, \overline{b}} \,
     \lambda^{\overline{b}}_- 
   + i
     \tilde{\alpha}_-
     h^{\overline{a} a} \,
     F_a                 
\end{aligned}
\label{SUSY}
\end{equation}
up to a total derivative.
In section
\ref{section:Supersymmetry-holomorphic},
we will demonstrate this invariance for the case in which the superpotential is holomorphic.
In section
\ref{section:Supersymmetry-nonholomorphic},
we will extend the analysis to the case in which the superpotential is not holomorphic.
There, we will find that supersymmetry imposes a constraint which relates the nonholomorphic parameters of the superpotential to the Hermitian curvature.
We will also find another constraint imposed by supersymmetry.
\section{\label{section:Supersymmetry-holomorphic}Supersymmetry invariance for holomorphic superpotential} 

In this section, we will show that, when the superpotential is holomorphic, the action
\eqref{action}
is invariant on-shell under the supersymmetry transformations
\eqref{SUSY}
up to a total derivative.
For these purposes, it is sufficient to set
$ \tilde{\alpha}_- = 0 $,\footnote{
The calculations for the case in which
$ \alpha_- = 0 $ 
and 
$ \tilde{\alpha}_- \neq 0 $
are analogous to those we will perform explicitly for the case in which
$ \alpha_- \neq 0 $ 
and 
$ \tilde{\alpha}_- = 0 $.
The general case, i.e.
$ \alpha_- $
and
$ \tilde{\alpha}_- $
both nonzero, is obtained by combining the above two cases. 
                                }
yielding 
\begin{equation}
\begin{aligned}
\delta \phi^i
  &= i 
     \alpha_-
     \psi^i_+ \, ,
\qquad      
&\delta \phi^{\overline{\imath}}
  &= 0 \, ,      
\\
\delta \psi^i_+ 
  &= 0 \, ,
\qquad  
&\delta \psi^{\overline{\imath}}_+
  &=
   - \alpha_-
     \partial_z \phi^{\overline{\imath}} \, ,
\\
\delta \lambda^{a}_-
  &= 
   - i
     \alpha_-
     \psi^j_+ \,
     A^a_{j b} \,
     \lambda^{b}_- 
   + i
     \alpha_-
     h^{a \overline{a}} \,
     \overline{F}_{\overline{a}} \, ,
\qquad
&\delta \lambda^{\overline{a}}_-
  &= 0 \, .         
\end{aligned}
\end{equation}
With this simplification, using the 
$ \lambda^a_- $ 
equation of motion,\footnote{This is valid because we have integrated out the auxiliary fields $ H^a $.}
we will show in section
\ref{subsection:BRST-exact-nonexact}
that the action
\eqref{action} 
can be written
\begin{align}
\label{action:Q-exact+non-exact}
S &= it \int_{\Sigma} d^2 z \,
     \left\{
             Q, V
     \right\}
   + t \int_{\Sigma}
      \Phi^*(K)
   + 2 t 
     \int_{\Sigma} d^2 z
     \left(
            \psi_+^{\overline{\imath}}
            \lambda_-^{\overline{a}} 
            \overline{D}_{\overline{\imath}} \overline{F}_{\overline{a}}
          - \psi_+^i 
            \lambda_-^{a} 
            D_a F_a
     \right),               
\end{align}
where
$ Q $
is the BRST operator, 
$ d^2 z = -i \, dz \wedge d{\overline{z}} $, 
\begin{equation}
V = 2 \left(
               g_{\overline{\imath} i}
               \psi^{\overline{\imath}}_+ 
               \overline{\partial}_{\overline{z}} \phi^i
             + i 
               \lambda^a_- 
               F_a         
      \right),
\end{equation}
and 
\begin{equation}
\int_{\Sigma} \Phi^*(K) 
   = \int_{\Sigma} d^2 z 
     \left(
            g_{i \overline{\imath}}
          + i 
            B_{i \overline{\imath}}  
     \right)
     \left(
            \partial_z \phi^i \,
            \overline{\partial}_{\overline{z}} \phi^{\overline{\imath}}
          - \overline{\partial}_{\overline{z}} \phi^i
            \partial_z \phi^{\overline{\imath}} 
     \right)          
\end{equation}
is the integral over the worldsheet 
$ \Sigma $
of the
pullback to
$ \Sigma $
of the complexified K{\"a}hler form
$ 
  K = 
    - i 
      \left(
             g_{i \overline{\imath}}
           + i
             B_{i \overline{\imath}} 
      \right)       
             d \phi^i
             \wedge
             d \phi^{\overline{\imath}}  
$.
Since 
$ \delta f = -i \alpha_- \{ Q, f \} $, 
where 
$ f $ 
is any field, the 
$ Q $-exact part of
\eqref{action:Q-exact+non-exact}
is
$ \delta $-exact 
and hence
$ \delta $-closed.  
In section
\ref{subsection:SUSY-invariance-topological},
we will complete our argument by establishing that the remaining terms are
$ \delta $-closed on shell up to a total derivative.
\subsection{\label{subsection:BRST-exact-nonexact}BRST-exact and non-exact terms}

Let us now derive
\eqref{action:Q-exact+non-exact}.
The BRST transformations are
\begin{equation}
\label{BRST}
\begin{aligned}
\left\{ 
        Q, \phi^i 
\right\}
  &= - \psi^i_+ \, ,
\qquad
&\left\{ 
         Q, \phi^{\overline{\imath}} 
 \right\} &= 0 \, ,
\\
\left\{ 
        Q, \psi^i_+ 
\right\} 
  &= 0 \, ,
\qquad
&\left\{ 
         Q, \psi^{\overline{\imath}}_+ 
 \right\} 
  &= - i \partial_z \phi^{\overline{\imath}} \, ,
\\
\left\{ 
        Q, \lambda^a_- 
\right\}
  &= \psi^j_+ 
     A^{a}_{j b} \lambda^b_- 
   - h^{a \overline{a}} \,
     \overline{F}_{\overline{a}} \, ,
\qquad
&\left\{ 
         Q, \lambda^{\overline{a}}_- 
 \right\}
  &= 0 \, .  
\end{aligned}
\end{equation}
Now, we compute
\begin{align*}
\frac{
       \left\{
               Q, V
       \right\} 
     }{2}
 &= \left\{
            Q,
            g_{\overline{\imath} i} 
            \psi^{\overline{\imath}}_+ 
            \overline{\partial}_{\overline{z}} 
            \phi^i
          + i   
            \lambda^a_- 
            F_a
    \right\}
\nonumber
\\[2ex]
  &= \left\{
             Q, g_{\overline{\imath} i}
     \right\}
     \psi^{\overline{\imath}}_+
     \overline{\partial}_{\overline{z}} 
     \phi^i
   + g_{\overline{\imath} i}
     \left\{
             Q, \psi^{\overline{\imath}}_+
     \right\}
     \overline{\partial}_{\overline{z}} 
     \phi^i
   - g_{\overline{\imath} i}
     \psi^{\overline{\imath}}_+
     \overline{\partial}_{\overline{z}}
     \left\{
             Q, \phi^i
     \right\}
\nonumber
\\[1ex]
  &\phantom{=} \
   + i
     \left\{
             Q, \lambda^a_-
     \right\}
     F_a
   - i
     \lambda^a_-
     \left\{
            Q, F_a
     \right\}
\nonumber
\\[2ex]
  &= \left(
            g_{\overline{\imath} i, k}
            \left\{
                    Q, \phi^k
            \right\}         
     \right)
     \psi^{\overline{\imath}}_+
     \overline{\partial}_{\overline{z}} 
     \phi^i
   + g_{\overline{\imath} i}
     \left(
          - i 
            \partial_z 
            \phi^{\overline{\imath}}
     \right)
     \overline{\partial}_{\overline{z}} 
     \phi^i
   - g_{\overline{\imath} i}
     \psi^{\overline{\imath}}_+
     \overline{\partial}_{\overline{z}}
     \left(
          - \psi^i_+
     \right)
\nonumber
\\[1ex]
  &\phantom{=} \
   + i
     \left(
            \psi^j_+ 
            A^a_{j b}
            \lambda^b_-
          - h^{a \overline{a}} \,
            \overline{F}_{\overline{a}}  
     \right)
     F_a
   - i 
     \lambda^a_-
     \left(
             F_{a, k}
             \left\{
                     Q, \phi^k
             \right\} 
     \right)
\allowdisplaybreaks
\nonumber
\\[2ex]
  &= g_{\overline{\imath} j}
     \Gamma^j_{i k}
     \left( 
          - \psi^k_+
     \right)       
     \psi^{\overline{\imath}}_+
     \overline{\partial}_{\overline{z}} 
     \phi^i
   - i 
     g_{\overline{\imath} i}
     \partial_z 
     \phi^{\overline{\imath}} \,
     \overline{\partial}_{\overline{z}} 
     \phi^i
   + g_{\overline{\imath} i}
     \psi^{\overline{\imath}}_+
     \overline{\partial}_{\overline{z}} 
     \psi^i_+
\nonumber
\\[1ex]
  &\phantom{=} \
   + i 
     \psi^j_+
     A^a_{j b}
     \lambda^b_-
     F_a 
   - i 
     h^{a \overline{a}} \,
     \overline{F}_{\overline{a}}
     F_a
   - i 
     \lambda^a_-
     F_{a, k}
     \left(
          - \psi^k_+
     \right)     
\allowdisplaybreaks
\nonumber
\\[2ex]
  &= 
    - \,
      i 
      g_{\overline{\imath} i}
      \partial_z 
      \phi^{\overline{\imath}} \,
      \overline{\partial}_{\overline{z}} 
      \phi^i
    + g_{\overline{\imath} i}
      \psi^{\overline{\imath}}_+
      \left(
             \overline{\partial}_{\overline{z}} 
             \psi^i_+
           + \overline{\partial}_{\overline{z}} 
             \phi^j \,
             \Gamma^i_{j k}
             \psi^k_+
      \right)
\nonumber
\\[1ex]
  &\phantom{=} \     
   - i 
     h^{a \overline{a}} \,
     \overline{F}_{\overline{a}}
     F_a
   - i 
     \psi^i_+
     \lambda^a_-
     \left(
            \partial_i 
            F_a
          - A^b_{i a}
            F_b  
     \right)
\nonumber
\\[2ex]
  &= 
   - \,
     i 
     g_{\overline{\imath} i}
     \partial_z 
     \phi^{\overline{\imath}} \,
     \overline{\partial}_{\overline{z}}
     \phi^i
   + g_{\overline{\imath} i}
     \psi^{\overline{\imath}}_+
     \overline{D}_{\overline{z}} 
     \psi^i_+
   - i 
     h^{a \overline{a}} \,
     \overline{F}_{\overline{a}}
     F_a
   - i 
     \psi^i_+
     \lambda^a_-
     D_i 
     F_a \, ,                                                                              
\end{align*}
where we have used
$ g_{\overline{\imath} i, k} = g_{\overline{\imath} j} \Gamma^{j}_{ik} $    
in the fourth step.
It follows that
\begin{equation*}
it 
\int_{\Sigma} 
d^2 z
\left\{
        Q, V
\right\}
   = 2t 
     \int_{\Sigma} 
     d^2 z
     \left(
            g_{\overline{\imath} i}
            \partial_z 
            \phi^{\overline{\imath}} \,
            \overline{\partial}_{\overline{z}} 
            \phi^i
          + i
            g_{\overline{\imath} i}
            \psi^{\overline{\imath}}_+
            \overline{D}_{\overline{z}} 
            \psi^i_+
          + h^{a \overline{a}} \,
            \overline{F}_{\overline{a}}
            F_a
          + \psi^i_+
            \lambda^a_-
            D_i 
            F_a      
     \right).
\end{equation*}
Using the identity
\begin{equation}
\int_{\Sigma} 
d^2 z \,
g_{\overline{\imath} i} 
\partial_z 
\phi^{\overline{\imath}} \,
\overline{\partial}_{\overline{z}} 
\phi^i
   = \int_{\Sigma} 
     d^2 z \,
     \frac{1}{2}
     \left(
            g_{\mu\nu}
          + i B_{\mu\nu} 
     \right)
     \partial_z \phi^{\mu} \,
     \partial_{\overline{z}} \phi^{\nu}
   - \frac{1}{2}
     \int_{\Sigma}
     \Phi^*(K) \, ,    
\end{equation}
we obtain
\begin{align*}
it \int_{\Sigma} d^2 z &
\left\{
        Q, V
\right\}
\nonumber
\\[1ex]
  &= 2 t 
     \int_{\Sigma} 
     d^2 z
     \left[
            \frac{1}{2}
            \left(
                   g_{\mu \nu}
                 + i 
                   B_{\mu \nu}
            \right)      
            \partial_z 
            \phi^{\mu}
            \overline{\partial}_{\overline{z}} 
            \phi^{\nu}
          + i
            g_{\overline{\imath} i} 
            \psi^{\overline{\imath}}_+
            \overline{D}_{\overline{z}} 
            \psi^i_+
          + h^{a \overline{a}} \,
            \overline{F}_{\overline{a}}
            F_a
          + \psi^i_+
            \lambda^a_-
            D_i 
            F_a      
     \right]
\nonumber
\\[1ex]
  &\phantom{=}     
   - \,
     t 
     \int_{\Sigma} 
     \Phi^*(K)
\nonumber
\\[1ex]
  &= S
   - t 
     \int_{\Sigma} 
     \Phi^*(K)
   - 2t 
     \int_{\Sigma} 
     d^2 z
     \left(
            i
            h_{\overline{a} a} 
            \lambda^a_-
            D_z 
            \lambda^{\overline{a}}_-
          + F_{i \overline{\imath} a \overline{a}} \,
            \psi^i_+
            \psi^{\overline{\imath}}_+
            \lambda^a_-
            \lambda^{\overline{a}}_-     
          + \psi^{\overline{\imath}}_+
            \lambda^{\overline{a}}_-
            \overline{D}_{\overline{\imath}} 
            \overline{F}_{\overline{a}}             
     \right),               
\end{align*}
and hence
\begin{align*}
S &= it 
     \int_{\Sigma} 
     d^2 z
     \left\{
             Q,         
             V
     \right\}   
   + t 
     \int_{\Sigma} 
     \Phi^*(K)
\nonumber
\\[1ex]
  &\phantom{=}     
   + \,
     2t 
     \int_{\Sigma} 
     d^2 z
     \left(
            i 
            h_{\overline{a} a}
            \lambda^a_-
            D_z 
            \lambda^{\overline{a}}_-
          + F_{i \overline{\imath} a \overline{a}} \,
            \psi^i_+
            \psi^{\overline{\imath}}_+
            \lambda^a_-
            \lambda^{\overline{a}}_-     
          + \psi^{\overline{\imath}}_+
            \lambda^{\overline{a}}_-
            D_{\overline{\imath}} 
            \overline{F}_{\overline{a}}             
     \right).
\end{align*}
An analogous result was found in \cite{MelnikovSethi:Half-twisted} for the case in which the gauge fields are absent and 
$ B = 0 $.
Finally, using the $ \lambda^a_- $ equation of motion
\begin{equation}
\label{eom:lambda^a_-}
\lambda^a_- :
\quad
  i
  h_{a \overline{a}}
  D_z \lambda^{\overline{a}}_-
+ F_{i \overline{\imath} a \overline{a}} \,
  \psi^i_+
  \psi^{\overline{\imath}}_+
  \lambda^{\overline{a}}_-
- \psi^i_+
  D_i
  F_a
= 0 \, ,  
\end{equation}
we obtain
\begin{equation*}
  i
  h_{a \overline{a}}
  \lambda^a_-
  D_z \lambda^{\overline{a}}_-
+ F_{i \overline{\imath} a \overline{a}} \,
  \psi^i_+
  \psi^{\overline{\imath}}_+
  \lambda^a_-
  \lambda^{\overline{a}}_-
=  
- \,
  \psi^i_+
  \lambda^a_-
  D_i F_a      
\end{equation*}
and hence
\begin{align*}
S &= it \int_{\Sigma} d^2 z \,
     \left\{
             Q, V
     \right\}
   + t \int_{\Sigma}
      \Phi^*(K)
   + 2 t 
     \int_{\Sigma} d^2 z
     \left(
            \psi_+^{\overline{\imath}}
            \lambda_-^{\overline{a}} 
            \overline{D}_{\overline{\imath}} \overline{F}_{\overline{a}}
          - \psi_+^i 
            \lambda_-^{a} 
            D_i F_a
     \right),               
\end{align*}
which is
\eqref{action:Q-exact+non-exact}.
\subsection{\label{subsection:SUSY-invariance-topological}Supersymmetry invariance of non-exact terms}

Let us now complete our argument that the action 
\eqref{action:Q-exact+non-exact}
is
$ \delta $-closed 
on shell up to a total derivative.
As we previously noted, the 
$ Q $-exact part of
\eqref{action:Q-exact+non-exact}
is
$ \delta $-exact
and hence 
$ \delta $-closed.
For the non-exact term of
\eqref{action:Q-exact+non-exact}
involving
$ \Phi^*(K) $,
note that
\begin{equation*}
\int_{\Sigma}
\Phi^*(K)
  = \int_{\Phi(\Sigma)} K
  = \int_{\Phi(\Sigma)}
    \left[
         - i 
           \left(
                  g_{i \overline{\imath}}
                + i
                  B_{i \overline{\imath}} 
           \right)
    \right]       
    d \phi^i
    \wedge
    d \phi^{\overline{\imath}}
\end{equation*}
and
$ K $
satisfies
\begin{equation*}
\partial K
 =
 - i \, 
   \partial_k
   \left(
          g_{i \overline{\imath}}
        + i
          B_{i \overline{\imath}} 
   \right)
    d \phi^k
    \wedge
    d \phi^i
    \wedge
    d \phi^{\overline{\imath}}
 = 0 \, .    
\end{equation*}
Thus,
\begin{equation}
\delta
\left[ 
       \Phi^*(K)
\right]       
  = \left[
           \Phi^*(K)
    \right]_k
    \delta
    \phi^k
  = 0 \, .         
\end{equation}
It remains to consider the non-exact expression of
\eqref{action:Q-exact+non-exact}
involving
$
  \psi_+^{\overline{\imath}}
  \lambda_-^{\overline{a}} 
  \overline{D}_{\overline{\imath}} \overline{F}_{\overline{a}}
- \psi_+^i 
  \lambda_-^{a} 
  D_i F_a
$.
First, we compute 
\begin{align}
\label{delta-non-exact-2a}
\delta
\left(
       \psi^{\overline{\imath}}_+
       \lambda^{\overline{a}}_- \,
       \overline{D}_{\overline{\imath}}
       \overline{F}_{\overline{a}}
\right)
  &= \left(
            \delta \psi_+^{\overline{\imath}} 
     \right)       
     \lambda^{\overline{a}}_- \,
     \overline{D}_{\overline{\imath}} \,
     \overline{F}_{\overline{a}}
   + \psi^{\overline{\imath}}_+
     \left(
            \delta \lambda^{\overline{a}}_-
     \right)       
     \overline{D}_{\overline{\imath}} \,
     \overline{F}_{\overline{a}}
   + \psi^{\overline{\imath}}_+
     \lambda^{\overline{a}}_i
     \left[
            \delta \!
            \left( \,
                   \overline{D}_{\overline{\imath}} \,
                   \overline{F}_{\overline{a}}
            \right) 
     \right]       
\allowdisplaybreaks
\nonumber
\\[1ex]
  &= \left(
          - \alpha_-
            \partial_z \phi^{\overline{\imath}}
     \right)       
     \lambda_-^{\overline{a}} \,
     \overline{D}_{\overline{\imath}} \,
     \overline{F}_{\overline{a}}
   + \psi^{\overline{\imath}}_+
     \lambda^{\overline{a}}_-
     \left[
            \delta
            \left(
                   \overline{\partial}_{\overline{\imath}}
                   \overline{F}_{\overline{a}}
                 - A^{\overline{b}}_{\overline{\imath} \, \overline{a}} \,
                   \overline{F}_{\overline{b}}   
            \right)
     \right]
\allowdisplaybreaks
\nonumber
\\[1ex]
  &= \left(
          - \alpha_-
            \partial_z \phi^{\overline{\imath}}
     \right)       
     \lambda_-^{\overline{a}} \,
     \overline{D}_{\overline{\imath}} \,
     \overline{F}_{\overline{a}}     
   + \psi^{\overline{\imath}}_+
     \lambda_-^{\overline{a}}
     \left[
            \overline{\partial}_{\overline{\imath}}
            \left( 
                   \delta \,
                   \overline{F}_{\overline{a}}
            \right)       
          - \left( 
                   \delta A^{\overline{b}}_{\overline{\imath} \, \overline{a}}
            \right)
            \overline{F}_{\overline{b}}
          - A^{\overline{b}}_{\overline{\imath} \, \overline{a}}
            \left(
                   \delta \,
                   \overline{F}_{\overline{b}} 
            \right)  
     \right]
\allowdisplaybreaks
\nonumber
\\[1ex]
  &= \left(
          - \alpha_-
            \partial_z \phi^{\overline{\imath}}
     \right)       
     \lambda_-^{\overline{a}} \,
     \overline{D}_{\overline{\imath}} \,
     \overline{F}_{\overline{a}}     
\allowdisplaybreaks
\nonumber
\\
  &\phantom{=}
   + \psi^{\overline{\imath}}_+
     \lambda_-^{\overline{a}}
     \left\{
            \overline{\partial}_{\overline{\imath}}
            \left[ \, 
                   \overline{F}_{\overline{a},k}
                   \left( 
                          \delta \phi^k
                   \right)       
            \right]       
          - \left[ 
                   A^{\overline{b}}_{\overline{\imath} \, \overline{a}, k}
                   \left( 
                          \delta \phi^k
                   \right)                   
            \right]
            \overline{F}_{\overline{b}}
          - A^{\overline{b}}_{\overline{\imath} \, \overline{a}}
            \left[
                   \overline{F}_{\overline{b}, k}
                   \left(
                          \delta \phi^k
                   \right) 
            \right]  
     \right\}
\allowdisplaybreaks
\nonumber
\\[1ex]
  &= \left(
          - \alpha_-
            \partial_z \phi^{\overline{\imath}}
     \right)       
     \lambda^{\overline{a}}_- \,
     \overline{D}_{\overline{\imath}} \,
     \overline{F}_{\overline{a}}     
   - \psi_+^{\overline{\imath}}
     \lambda_-^{\overline{a}}
     A^{\overline{b}}_{\overline{\imath} \, \overline{a}, k}
     \left( 
            i
            \alpha_-
            \psi^k_+                   
     \right)                   
    \overline{F}_{\overline{b}} \, ,                         
\end{align}
where we have used
$ \overline{F}_{\overline{a},k}  = 0 $
in the last step.
Now, we compute
\begin{align}
\label{delta-non-exact-2b}
\delta
\left(
    - \psi^i_+ 
       \lambda^a_-
       D_i F_a
\right)
  &= 
   - \,
     \left(
            \delta \psi^i_+
     \right)       
     \lambda^a_- 
     D_i F_a
   - \psi^i_+
     \left(
            \delta \lambda^a_-
     \right)        
     D_i F_a
   - \psi^i_+
     \lambda^a_-
     \left[
            \delta \!
            \left( 
                   D_i F_a
            \right)
     \right]       
\nonumber
\\[1ex]
  &= 
   - \,
     \psi^i_+
     \left(
          - i
            \alpha_-
            \psi^j_+ \,
            A^a_{j b} \,
            \lambda^b_-     
          + i
            \alpha_-
            h^{a \overline{a}} \,
            \overline{F}_{\overline{a}}
     \right) 
     D_i F_a
   - \psi^i_+
     \lambda^a_-
     \left[
            \delta \!
            \left(
                   \partial_i F_a
                 - A^b_{i a}
                   F_b  
            \right)
     \right]
\nonumber
\\[1ex]
  &=
   - \,
     \psi^i_+
     \left(
          - i
            \alpha_-
            \psi^j_+ \,
            A^a_{j b} \,
            \lambda^b_-     
     \right) 
     \Bigl(
            \partial_i F_a - A^c_{i a} F_c
     \Bigr)
\nonumber
\\
  &\phantom{=} \      
   - \,
     \psi_+^i
     \left(     
            i
            \alpha_-
            h^{a \overline{a}} \,
            \overline{F}_{\overline{a}}
     \right) 
     D_i F_a               
   - \psi^a_+
     \lambda^a_-
     \left[
            \partial_i 
            \left( 
                   \delta F_a
            \right)       
          - \left( 
                   \delta A^b_{i a}
            \right)       
            F_b  
          - A^b_{i a}
            \left(
                   \delta F_b
            \right)
     \right]
\nonumber
\\[1ex]
  &= - \,
     \psi^i_+
     \left(
          - i
            \alpha_-
            \psi^j_+ \,
            A^a_{j b} \,
            \lambda^b_-     
     \right) 
     \partial_i F_a
\nonumber
\\
  &\phantom{=} \
   - \psi^i_+
     \left[
          - i
            \alpha_-
            \psi^j_+
            \left(
                   h^{a \overline{b}} \,
                   \partial_j
                   h_{\overline{b} b}
            \right)
            \lambda^b_- 
     \right]
     \Bigl[
          - \left(
                   h^{c \overline{c}} \,
                   \partial_i
                   h_{\overline{c} a}
            \right)
            F_c 
     \Bigr]     
\nonumber
\\
  &\phantom{=} \
   + i
     \alpha_-
     h^{a \overline{b}} \,
     \overline{F}_{\overline{b}}
     \left(
            \psi^i_+
            D_i F_a
     \right)
\nonumber
\\
  &\phantom{=} \          
   - \psi^i_+
     \lambda^a_-
     \left\{
            \partial_i 
            \left[ 
                   F_{a,k}
                   \left(
                          \delta \phi^k
                   \right)
            \right]       
          - \left[ 
                   A^b_{i a, k}
                   \left(
                          \delta \phi^k
                   \right)
            \right]       
            F_b  
          - A^b_{i a}
            \left[
                   F_{b, k}
                   \left(
                          \delta \phi^k
                   \right)                   
            \right]
     \right\}
\allowdisplaybreaks
\nonumber
\\[1ex]
  &= - \,
     \psi^i_+
     \left(
          - i
            \alpha_-
            \psi^j_+ \,
            A^a_{j b} \,
            \lambda^b_-     
     \right) 
     \partial_i F_a
\nonumber
\\
  &\phantom{=} \
   - \psi^i_+
     \left[
          - i
            \alpha_-
            \psi^j_+
            \left(
                   h^{a \overline{b}} \,
                   \partial_j
                   h_{\overline{b} b}
            \right)
            \lambda^b_- 
     \right]
     \Bigl[
          - \left(
                 - h_{\overline{c} a} \,
                   \partial_i
                   h^{c \overline{c}}
            \right)
            F_c 
     \Bigr]     
\nonumber
\\
  &\phantom{=} \
   + i
     \alpha_-
     h^{a \overline{b}} \,
     \overline{F}_{\overline{b}}
     \left(
            i
            h_{a \overline{a}}
            D_z \lambda^{\overline{a}}_-
          + F_{i \overline{\imath} a \overline{a}} \,
            \psi^i_+
            \psi^{\overline{\imath}}_+
            \lambda^{\overline{a}}_-            
     \right)
\nonumber
\\
  &\phantom{=} \              
   - \psi^i_+
     \lambda^a_-
     \left[ 
            \partial_i
            F_{a, k}
            \left(
                   i
                   \alpha_-
                   \psi^k_+
            \right)      
          - A^b_{i a, k}
            \left(
                   i
                   \alpha_-
                   \psi^k_+
            \right)       
            F_b  
          - A^b_{i a}
            F_{b, k}
            \left(
                   i
                   \alpha_-
                   \psi^k_+
            \right)                   
     \right]
\allowdisplaybreaks
\nonumber     
\\[1ex]
  &= 
   - \,
     \psi^i_+
     \left(
          - i
            \alpha_-
            \psi^j_+ \,
            A^a_{j b} \,
            \lambda^b_-     
     \right) 
     \partial_i F_a
\nonumber
\\
  &\phantom{=} \
   - \psi^i_+
     \left[
          - i
            \alpha_-
            \psi^j_+
            \left(
                   \partial_i
                   h^{a \overline{b}} \,
                   \partial_j
                   h_{\overline{b} b}
            \right)
            \lambda^b_- 
            F_a 
     \right]     
\nonumber
\\
  &\phantom{=} \   
   - \alpha_-
     \overline{F}_{\overline{a}}
     D_z \lambda^{\overline{a}}_-
   + \left(
            i
            \alpha_-
            h^{a \overline{b}} \,
            \overline{F}_{\overline{b}}
     \right)    
     F_{i \overline{\imath} a \overline{a}} \,
     \psi^i_+
     \psi^{\overline{\imath}}_+
     \lambda^{\overline{a}}_-
\nonumber
\\
  &\phantom{=} \ 
   + \psi^i_+
     \left(
          - i
            \alpha_-
            \psi^j_+ \,
            A^a_{j b, i} \,
            \lambda^b_-
            F_a 
     \right)
   + \psi^i_+
     \left(
          - i
            \alpha_-
            \psi^j_+ \,
            A^a_{j b} \,
            \lambda^b_-     
     \right) 
     \partial_i F_a
\allowdisplaybreaks
\nonumber     
\\[1ex]
  &= 
   - \,
     \psi^i_+
     \left[
          - i
            \alpha_-
            \psi^j_+
            \left(
                   \partial_i
                   h^{a \overline{b}} \,
                   \partial_j
                   h_{\overline{b} b}
            \right)
            \lambda^b_- 
            F_a 
     \right]
\nonumber
\\
  &\phantom{=} \   
   - \alpha_-
     \overline{F}_{\overline{a}}
     D_z \lambda^{\overline{a}}_-
   + \left(
            i
            \alpha_-
            h^{a \overline{b}} \,
            \overline{F}_{\overline{b}}
     \right)    
     F_{i \overline{\imath} a \overline{a}} \,
     \psi^i_+
     \psi^{\overline{\imath}}_+
     \lambda^{\overline{a}}_-
\nonumber
\\
  &\phantom{=} \ 
   + \psi^i_+
     \left[
          - i
            \alpha_-
            \psi^j_+
            \left(
                   \partial_i
                   h^{a \overline{b}} \,
                   \partial_j
                   h_{\overline{b} b}
                 + h^{a \overline{b}} \,
                   \partial_i
                   \partial_j
                   h_{\overline{b} b}
            \right)
            \lambda^b_-
            F_a 
     \right]
\allowdisplaybreaks
\nonumber     
\\[1ex]
  &=
   - \,
     \alpha_-
     \overline{F}_{\overline{a}}
     D_z \lambda^{\overline{a}}_-
   + \left(
            i
            \alpha_-
            h^{a \overline{b}} \,
            \overline{F}_{\overline{b}}
     \right)    
     F_{i \overline{\imath} a \overline{a}} \,
     \psi^i_+
     \psi^{\overline{\imath}}_+
     \lambda^{\overline{a}}_- \ .                                                                        
\end{align}
where we have used
$ 
  A^a_{j b} 
= h^{a \overline{b}} \,
  \partial_j
  h_{\overline{b} b}
$
in the fourth step, the identity
$
  h^{c \overline{c}} \,
                   \partial_i
                   h_{\overline{c} a}
= - h_{\overline{c} a} \,
                   \partial_i
                   h^{c \overline{c}}                   
$
and the 
$ \lambda^a_- $
equation of motion
\eqref{eom:lambda^a_-}
in the fifth step,
$
  \psi^i_+
  \lambda^a_- \,
  \partial_i
   F_{a, k}
  \left(
         i
         \alpha_-
         \psi^k_+
  \right)
= 0            
$
in the sixth step,
$
  A^a_{j b, i}
= \partial_i
  h^{a \overline{b}} \,
  \partial_j
  h_{\overline{b} b}
+ h^{a \overline{b}} \,
  \partial_i
  \partial_j
  h_{\overline{b} b}  
$
in the seventh step, and
\begin{equation*}
  \psi^i_+
  \left[
         i
         \alpha_-
         \psi^j_+
         \left(
                h^{a \overline{b}} \,
                \partial_i
                \partial_j
                h_{\overline{b} b}
         \right)       
         \lambda^b_-
         F_a 
  \right]
= 0     
\end{equation*}
in the last step.
Note that the first term on the right-hand side of
\eqref{delta-non-exact-2b}
cancels the first term on the right-hand side of
\eqref{delta-non-exact-2a} up to a total derivative: 
\begin{align}
\label{cancel-first-terms-rhs-nonexact-2a-2b}
- \,
    \alpha_-
    \overline{F}_{\overline{a}} \,
    D_z \lambda^{\overline{a}}_-
 &=
  - \,
    \alpha_-
    \overline{F}_{\overline{a}}
    \left(
           \partial_z \lambda^{\overline{a}}_-
         + \partial_z \phi^{\overline{\imath}} \,
           A^{\overline{a}}_{\overline{\imath} \, \overline{b}} \,
           \lambda^{\overline{b}}_-
    \right)
\nonumber
\\[1ex]
  &= \alpha_-
     \left( 
            \overline{F}_{\overline{a}, k} \,
            \partial_z \phi^k
          + \overline{F}_{\overline{a}, \overline{k}} \,
            \partial_z \phi^{\overline{k}} 
     \right)
     \lambda^{\overline{a}}_-
   - \alpha_-
     \partial_z \!
     \left(
            \overline{F}_{\overline{a}} \,
            \lambda^{\overline{a}}_-            
     \right)
   - \alpha_-
     \overline{F}_{\overline{a}} \,
     \partial_z \phi^{\overline{\imath}} \,
     A^{\overline{a}}_{\overline{\imath} \, \overline{b}} \,
     \lambda^{\overline{b}}_- 
\nonumber
\\[1ex]
 &= \left(
           \alpha_-
           \partial_z \phi^{\overline{\imath}}
    \right)       
    \lambda^{\overline{a}}_-
    \left(
           \overline{\partial}_{\overline{\imath}}
           \overline{F}_{\overline{a}}
         - A^{\overline{b}}_{\overline{\imath} \, \overline{a}} \,
           \overline{F}_{\overline{b}}   
    \right)
  - \alpha_-
    \partial_z \!
    \left(
           \overline{F}_{\overline{a}} \,
           \lambda^{\overline{a}}_-
     \right)   
\nonumber
\\[1ex]
 &= \left(
           \alpha_-
           \partial_z \phi^{\overline{\imath}}
     \right)       
     \lambda^{\overline{a}}_- \,
     \overline{D}_{\overline{\imath}} \,
     \overline{F}_{\overline{a}}
   - \alpha_-
     \partial_z \!
     \left(
            \overline{F}_{\overline{a}} \,
            \lambda^{\overline{a}}_-
     \right),                            
\end{align}
where we used
$ \overline{F}_{\overline{a}, k} = 0 $
in the fourth step.
Furthermore, the second term on the right-hand side of
\eqref{delta-non-exact-2b}
cancels the second term on the right-hand side of
\eqref{delta-non-exact-2a}:
\begin{align}
\left(
       i
       \alpha_-
       h^{a \overline{b}} \,
       \overline{F}_{\overline{b}}
\right)    
F_{i \overline{\imath} a \overline{a}} \,
\psi^i_+
\psi^{\overline{\imath}}_+
\lambda^{\overline{a}}_-
 &= \left(
           i
           \alpha_-
           h^{a \overline{c}} \,
           \overline{F}_{\overline{c}}
    \right)
    \left(
           h_{a \overline{b}} \,
           A^{\overline{b}}_{\overline{\imath} \, \overline{a}, i}
    \right)
    \psi^{\overline{\imath}}_+
    \lambda^{\overline{a}}_-
    \psi^i_+
\nonumber
\\[1ex]
 &= \psi^{\overline{\imath}}_+
    \lambda^{\overline{a}}_- \,
    A^{\overline{b}}_{\overline{\imath} \, \overline{a}, k} \,
    \left(
           i
           \alpha_-
           \psi^k_+
    \right)
    \overline{F}_{\overline{b}} \, ,           
\end{align}
where we have used
$
  F_{i \overline{\imath} a \overline{a}}
= h_{a \overline{b}} \,
  A^{\overline{b}}_{\overline{\imath} \, \overline{a}, i}  
$
in the first step.
It follows that
\eqref{delta-non-exact-2b}
cancels
\eqref{delta-non-exact-2a}
up to a total derivative, i.e.
\begin{equation}
\delta
\left(
    - \psi^i_+ 
       \lambda^a_-
       D_i F_a
\right)
  = 
  - \,
    \delta
    \left(
           \psi^{\overline{\imath}}_+
           \lambda^{\overline{a}}_- \,
           \overline{D}_{\overline{\imath}}
           \overline{F}_{\overline{a}}
    \right)
  - \alpha_-
    \partial_z \!
    \left(
           \overline{F}_{\overline{a}} \,
            \lambda^{\overline{a}}_-
    \right).  
\end{equation}
This completes our argument.
\section{\label{section:Supersymmetry-nonholomorphic} Supersymmetry invariance for nonholomorphic superpotential}

In this section, we will extend the analysis of section
\ref{section:Supersymmetry-holomorphic} 
to the case in which the superpotential is not holomorphic.
This requires revisting the steps in
\eqref{delta-non-exact-2a}
and
\eqref{cancel-first-terms-rhs-nonexact-2a-2b}
where we used
$ \overline{F}_{\overline{a}, k} = 0 $.
Allowing for
$ \overline{F}_{\overline{a}, k} \ne 0 $,
\eqref{cancel-first-terms-rhs-nonexact-2a-2b}
becomes    
\begin{equation*}
- \,
  \alpha_-
  \overline{F}_{\overline{a}} \,
  D_z \lambda^{\overline{a}}_-
= \left(
           \alpha_-
           \partial_z \phi^{\overline{\imath}}
     \right)       
     \lambda^{\overline{a}}_- \,
     \overline{D}_{\overline{\imath}} \,
     \overline{F}_{\overline{a}}
   - \alpha_-
     \partial_z \!
     \left(
            \overline{F}_{\overline{a}} \,
            \lambda^{\overline{a}}_-
     \right)
   + \alpha_- 
     \overline{F}_{\overline{a}, k} \,
     \partial_z \phi^k
     \lambda^{\overline{a}}_- \, .    
\end{equation*}
It follows that the cancellation described by
\eqref{cancel-first-terms-rhs-nonexact-2a-2b}
will still apply provided that the constraint
\begin{equation}
\label{constraint-1}
\overline{F}_{\overline{a}, k} \,
\partial_z \phi^k
\lambda^{\overline{a}}_-
  = 0     
\end{equation}
is satisfied.
Furthermore, in the next to last line of
\eqref{delta-non-exact-2a}, 
we now have
\begin{align}
\psi^{\overline{\imath}}_+
\lambda_-^{\overline{a}}
\left\{
        \overline{\partial}_{\overline{\imath}}
        \left[ \, 
               \overline{F}_{\overline{a},k}
               \left( 
                      \delta \phi^k
               \right)       
        \right]       
\right.
     &- A^{\overline{b}}_{\overline{\imath} \, \overline{a}}
\left.
        \left[
               \overline{F}_{\overline{b}, k}
               \left(
                      \delta \phi^k
               \right) 
        \right]  
\right\}
\nonumber
\\[1ex]
 &= \psi^{\overline{\imath}}_+
    \lambda_-^{\overline{a}}
    \left\{
           \overline{\partial}_{\overline{\imath}}
           \left[ 
                  \overline{F}_{\overline{a},k}
                  \left(
                         i
                         \alpha_-
                         \psi^k_+
                   \right)
           \right]                     
         - A^{\overline{b}}_{\overline{\imath} \, \overline{a}} \,
           \overline{F}_{\overline{b}, k}
           \left(
                  i
                  \alpha_-
                  \psi^k_+
           \right)  
    \right\}
\nonumber
\\[1ex]
 &= \psi^{\overline{\imath}}_+
    \lambda_-^{\overline{a}}
    \left(
           \overline{\partial}_{\overline{\imath}} 
           \overline{F}_{\overline{a},i}              
         - A^{\overline{b}}_{\overline{\imath} \, \overline{a}} \,
           \overline{F}_{\overline{b}, i}  
    \right)
    \left(
           i
           \alpha_-
           \psi^i_+
    \right)
\nonumber
\\[1ex]
 &= \psi^{\overline{\imath}}_+
    \lambda_-^{\overline{a}}
    \left[
           \overline{\partial}_{\overline{\imath}} 
           \overline{F}_{\overline{a}, i}              
         + A^{\overline{b}}_{\overline{\imath} \, \overline{a}, i} \,
           \overline{F}_{\overline{b}}
         - \partial_i \!
           \left(
                  A^{\overline{b}}_{\overline{\imath} \, \overline{a}} \,
                  \overline{F}_{\overline{b}}
           \right)     
    \right]
    \left(
           i
           \alpha_-
           \psi^i_+
    \right)
\nonumber
\\[1ex]
 &= \psi^{\overline{\imath}}_+
    \lambda_-^{\overline{a}}
    \left[
           \partial_i \!
           \left(
                  \overline{\partial}_{\overline{\imath}} 
                  \overline{F}_{\overline{a}}
                - A^{\overline{b}}_{\overline{\imath} \, \overline{a}} \,
                  \overline{F}_{\overline{b}}  
           \right)              
         + A^{\overline{b}}_{\overline{\imath} \, \overline{a}, i} \,
           \overline{F}_{\overline{b}}  
    \right]
    \left(
           i
           \alpha_-
           \psi^i_+
    \right)
\nonumber
\\[1ex]
 &= \psi^{\overline{\imath}}_+
    \lambda_-^{\overline{a}}
    \left(
           \partial_i
           \overline{D}_{\overline{\imath}} 
           \overline{F}_{\overline{a}}              
         + F_{i \overline{\imath} \, a \overline{a}} \,
           h^{a \overline{b}} \,
           \overline{F}_{\overline{b}}  
    \right)
    \left(
           i
           \alpha_-
           \psi^i_+
    \right),            
\end{align}
where we have used 
$ 
  A^{\overline{b}}_{\overline{\imath} \, \overline{a}, i}
= h^{a \overline{b}} \,
  F_{i \overline{\imath} \, a \overline{a}}
$  
in the last step.
It follows that, in addition to requiring
\eqref{constraint-1},
supersymmetry imposes the constraint
\begin{equation}
\label{constraint-2}
  \partial_i
  \overline{D}_{\overline{\imath}} 
  \overline{F}_{\overline{a}}              
+ F_{i \overline{\imath} \, a \overline{a}} \,
  h^{a \overline{b}} \,
  \overline{F}_{\overline{b}}
= 0 \, .             
\end{equation}
Various special cases of
\eqref{constraint-2}
were used in
\cite{GaravusoSharpe:Analogues}
to establish properties of Mathai-Quillen form analogues which arise in the corresponding heterotic Landau-Ginzburg models.
In that paper, it was claimed that supersymmetry imposes those constraints.
In this paper, we have worked out the details supporting that claim.     


\end{document}